\documentstyle[emulateapj,psfig,apjfonts]{article}
\submitted{{\it Accepted for publication in AJ}}
\begin{document}

\title{Spectroscopic Properties of Cool Stars in the SDSS: An Analysis of
Magnetic Activity and a Search for Subdwarfs}

\author{Andrew A. West\altaffilmark{1,2}, 
Suzanne L. Hawley\altaffilmark{2}, 
Lucianne M. Walkowicz\altaffilmark{2},
Kevin R. Covey\altaffilmark{2},
Nicole M. Silvestri\altaffilmark{2},
Sean N. Raymond\altaffilmark{2},
Hugh C. Harris\altaffilmark{3},
Jeffrey A. Munn\altaffilmark{3},
Peregrine M. McGehee\altaffilmark{4,5},
\v{Z}eljko Ivezi\'{c}\altaffilmark{6,7},
J. Brinkmann\altaffilmark{8}}

\altaffiltext{1}{Corresponding author: west@astro.washington.edu}
\altaffiltext{2}{Department of Astronomy, University of Washington, Box 351580,
Seattle, WA 98195}
\altaffiltext{3}{US Naval Observatory, Flagstaff Station, P.O. Box 1149, Flagstaff, AZ 86002}
\altaffiltext{4}{NMSU, Department of Astronomy, 1320 Frenger Mall, Las Cruces, NM 88003}
\altaffiltext{5}{Los Alamos National Laboratory, MS H820, Los Alamos, NM 87545}
\altaffiltext{6}{Princeton University, Princeton, NJ 08544}
\altaffiltext{7}{H.N. Russell Fellow, on leave from the University of
Washington}
\altaffiltext{8}{Apache Point Observatory, P.O. Box 59, Sunspot, NM 88349}

%need right affiliations for USNO and PMcG

\begin{abstract} We present a spectroscopic analysis of nearly 8000
late-type dwarfs in the Sloan Digital Sky Survey.  Using the H$\alpha$
emission line as an activity indicator, we investigate the fraction of
active stars as a function of spectral type and find a peak near type M8,
confirming previous results.  In contrast to past findings, we find
that not all M7-M8 stars are active.  We show that this may be a
selection effect of the distance distributions of previous samples,
as the active stars appear to be concentrated near the Galactic Plane.
We also examine the activity strength (ratio of the
luminosity emitted in H$\alpha$ to the bolometric luminosity) for
each star, and find that the mean activity strength is constant over the
range M0-M5 and declines at later types.  The decline begins
at a slightly earlier spectral type than 
previously found.  We explore the effect that activity has on the broadband
photometric colors and find no significant differences between active and 
inactive stars.  
We also carry out a search 
for subdwarfs using spectroscopic metallicity indicators, and find 60 subdwarf 
candidates. Several of these candidates are near the extreme subdwarf 
boundary.  The spectroscopic subdwarf candidates are redder by 
$\sim 0.2$ magnitudes in $g-r$ compared to disk dwarfs at the same $r-i$ color.

\end{abstract}


\section{Introduction}
Low mass stars with late spectral types (M,L) are the majority 
constituent of the Galaxy by number.  Their main-sequence lifetimes 
are much greater than the current age of the Universe and they therefore serve
as useful probes of Galactic star formation history in the local
solar neighborhood (Gizis, Reid \& Hawley 2002).
They also encompass many important regions of stellar parameter space,
including the onset of complete convection in the stellar interior, the 
onset of significant electron degeneracy in the core, and the formation 
of dust and subsequent depletion of metals onto dust grains in the stellar
atmosphere.  Of particular interest is the fact that many late-type stars
have strong surface magnetic fields (Johns-Krull \& Valenti 1996) that 
heat the outer atmosphere above the photosphere, and lead to observable
emission from the chromosphere (e.g. Ca II and H Balmer series lines), 
the transition region (e.g resonance lines of abundant ions such as C IV),
and the corona (e.g. thermal soft X-rays).  The physics that controls 
the production of magnetic fields in low mass stars is not well understood, 
as the lack of a radiative-convective boundary layer precludes storing 
large scale fields as in a solar-type dynamo.  However,
recent work on turbulent dynamo mechanisms (Bercik et al. 2004) may soon 
provide new insight into the formation and properties of magnetic fields in
low mass stars.  The present study seeks to place strong empirical
constraints on the magnetic activity in these stars, by measuring optical 
H$\alpha$ emission in a sample of nearly 8000 low mass dwarfs.
Our results will aid in the interpretation of the models and provide a 
connection to the physical processes occurring in the atmospheres of
the stars.

The H$\alpha$ emission line is produced by collisional excitation in the
relatively dense chromospheres of these low mass, high gravity stars.  
Located in the red region of the optical spectrum, it is the strongest and 
best-studied indicator of magnetic activity in late type stars (in contrast
to solar type stars which are usually studied in the Ca II H and K 
resonance lines).  Previous results indicate that the
fraction of M dwarfs with H$\alpha$ emission increases from early to mid M
spectral types (Joy \& Abt 1974; Hawley, Gizis \& Reid 1996), reaches a peak near type M7 
where essentially all stars are active, and declines toward later types (Gizis
et al. 2000a).  The increasing fraction in the early to mid M types may
be an age effect such that the activity lasts longer in the mid M types
(Hawley, Reid \& Tourtellot 2000a; Gizis et al. 2002), while the decline toward late M types
may reflect difficulty in producing and/or maintaining surface magnetic fields
due to the physics of turbulent dynamos and/or the increasingly neutral 
atmosphere (Hawley, Reid \& Gizis 2000b; Fleming, Giampapa \& Schmitt 2000;
Mohanty et al. 2002).  It does not appear that
the atmosphere changes character such that H$\alpha$ is no longer produced
when magnetic activity is present in the later type stars.  
Indeed the ratio of H$\alpha$ to transition
region (C IV) and soft X-ray emission appears to be similar throughout the
M dwarf spectral sequence (Hawley \& Johns-Krull 2003).

Previous work also shows that the activity strength, measured by the ratio 
of the luminosity in H$\alpha$ to the bolometric luminosity 
(L$_{\rm{H}_\alpha}$/L$_{bol}$), is nearly constant (with large scatter)
through the M0-M6 range (Hawley et al. 1996) and declines at later
types (Burgasser et al. 2002, Cruz \& Reid 2002).
However, the sample of stars on which these conclusions are based numbers
less than 100 for types M7 and later, due to the difficulty of obtaining
spectra for these faint objects.  Our sample in this crucial spectral
type range is an order of magnitude larger, comprising more than 1000 stars 
with types between M7 and L0, and nearly 8000 stars between M0 and L0.
This uniformly acquired and reduced sample from the Sloan Digital
Sky Survey (SDSS), with well-characterized uncertainties, now allows us 
to obtain statistically significant magnetic activity results for the entire 
M dwarf spectral type range.  In addition, we compare the photometric colors
of the active and inactive stars to investigate whether the active
stars are typically bluer at the same spectral type, an effect attributed
to an increase in plage areas on active stars by
Amado \& Byrne (1997).  Alternatively, such a correlation might indicate that 
micro-flaring activity plays a significant role in heating the outer
atmospheres of active M dwarfs (cf. G\"{u}del et al. 2003).

Our SDSS spectral sample is also well-suited for investigation of 
the incidence of
low mass subdwarfs in the local neighborhood.  Gizis (1997) showed
that M subdwarfs could be identified by comparing the metallicity and 
gravity sensitive 
molecular bands of CaH and TiO.  Using his technique, we identify 
dozens of new M subdwarf and extreme subdwarf candidates, and examine
their colors in the SDSS $ugriz$ filters.

\section{Data} The SDSS (York et al. 2000; Gunn et al. 1998; Fukugita et al. 1996; Hogg et al. 2001; Smith et al. 2002; Stoughton et al. 2002; Pier et al. 2003, Abazajian et al. 2003)
is an excellent tool for spectroscopic studies of stars.  Each spectroscopic
plate produces 640 spectra with resolution $\lambda/{\Delta\lambda} = 1800$.
Because all spectra are obtained and reduced
through uniform pipelines with well understood uncertainties (York et al. 2000), 
it is possible to undertake statistically robust studies of very large
samples.  Our sample was selected from the SDSS spectroscopic and photometric
databases as follows.  In July 2002, the spectroscopic database was
queried for all objects which satisfied the SDSS color cuts for 
late-type stars given in Hawley et al. (2002; hereafter H02), viz.
$0.8 < (r-i) < 3.0$ and $0.3 < (i-z) < 3.5$.  The photometry for the
color cuts came from the preliminary SDSS photometric 
database\footnote{the ``Chunk'' database} available at that time.
This color cut yielded $\sim17000$ late-type star candidates with SDSS 
spectra.  In April 2003 we added an additional $\sim 5000$ candidates from
spectroscopic observations that were specifically targeted for late type
stars in the Southern Survey region.  The final $\sim22000$ candidates
come from 427 SDSS spectroscopic plates; 380 of the plates are in the SDSS 
Data Release 2 (DR2; Abazajian et al. 2004) sky area while
47 plates have yet to be publicly released.  
%All of the spectra were reduced through rerun 15 or 20 of the SDSS spectroscopic pipeline, which wereavailable to us at the time the sample was defined.  Because
Due to recent improvements in the photometric reductions, we have used 
photometry only from the 5\_4\_25 version of Photo 
(SDSS photometric reduction software) in our final
analysis.  This is the same version used in the DR2 release.
\section{Analysis}

\subsection{Spectral Types} 

In order to assign spectral types to the stars
in the sample, we pass each spectrum through two independent spectral
typing pipelines.  The first method determines the best fit, in a least
squares sense, to both the SDSS late-type template spectra defined in H02 
and the template spectra from the Keck and Apache Point Observatories
used by H02 in their analysis; it is
a derivative of the procedure originally described in Kirkpatrick (1991)
and subsequently modified by Henry et al. (2002).  
The second method measures several molecular 
band indices in each spectrum and uses a weighted mean of the individual 
types from each index to assign a best-guess spectral type (cf. Reid, 
Hawley, \& Gizis 1995a; Kirkpatrick et al. 1999).  Both methods are described in 
detail in H02.  
Using the spectral typing pipeline outputs as a guide, each spectrum
is individually inspected by eye and the assigned spectral type is altered 
as needed.  As reported in H02, the uncertainty in this procedure is $\sim$ 1 spectral type.
Spectra with a signal-to-noise ratio (SNR) $<$ 5 are not typed, as
experience shows that they do not give useful results in
our analysis.  Of the 22000 original candidates, $\sim9200$ were assigned
M or L spectral types and had spectra with the necessary signal-to-noise ratio
for further analysis.  An important byproduct of the spectral typing procedure
was the identification of numerous white dwarf - M dwarf binary systems
in our sample.  We find the binaries based on the appearance of blue 
continuum flux and features such as broad hydrogen Balmer absorption lines
(Raymond et al. 2003).  We have removed all such 
binaries from the final sample so that their colors will not affect our 
investigation of the colors of active vs. inactive M dwarfs, and because 
the nature of the M dwarf activity may be influenced by the presence
of a close binary companion.  $\S$3.3 outlines a photometric test used to further constrain our sample to stars without white dwarf companions.  Our sample is thus, as far as possible, composed
of only single M dwarfs (or M dwarfs in binary systems with fainter M or 
L dwarfs that we are unable to discern from the spectra).

\subsection{Activity} 

\subsubsection{$\rm{H}\alpha$ Measurements} 
The SDSS spectra are spectrophotometrically calibrated,
so we measure the H$\alpha$ flux directly from each spectrum.  The challenge
is to automatically, but accurately, measure the flux in
H$\alpha$ over a range of spectral types with changing continuum
slope and molecular features.  Our procedure uses
trapezoidal integration to sum the flux under the emission line, taking
care to avoid and/or account for regions that are affected by local TiO 
features including the continuum bump just to the blue of H$\alpha$  
(Pettersen \& Coleman 1981) and the absorption trough to the red
(Tinney \& Reid 1998; White \& Basri 2003).  We define a
14\AA\ interval (7\AA\ on either side of the line center) for the H$\alpha$
flux calculation.   We investigated several options to find the optimum
continuum flux measure: a) the average over two 50\AA\  sections 
(6500-6550\AA\ and 6575-6625\AA)  on either side of the line;
b) the average over two 10\AA\ sections (6545-6555\AA\ and 6575-6585\AA)
on either side of the line; and c) the blue 10\AA\ region only, as used
by Reid et al.(1995a) for the PMSU survey (6545-6555\AA).  We found no
significant differences in the mean continuum values for these three
measures, and therefore chose to use measure (a) as it is the most
robust, and because we also use the 50\AA\ regions as comparison for the equivalent width (EW) measurement (see below).
The formal uncertainty in the flux at each wavelength is available
from the SDSS spectral calibration and is propagated at every stage 
of the analysis.

In order to be defined as active, each spectrum must pass the following
criteria: 1) the H$\alpha$ EW must be greater than 1.0\AA; 2) the uncertainty
in the EW must be smaller than the EW value;  3) the height of the
H$\alpha$ emission line must be 3 times the noise at line center; and
4) the EW of H$\alpha$ must be 3 times larger than the average EW of the 50\AA\ comparison regions.  
We examined subsets of the spectra that pass
less conservative criteria, including those that passed only 2 or 3 of
the 4 tests.  These account for less than 5\% of the total
sample, and upon inspection are clearly marginal.  There are 430 stars
in the less conservative subsets that appear to be active
but have H$\alpha$ EW less than 1.0\AA.  We do not include these in 
our analysis but mention them as potential active stars similar to the 
stars in Groups C and E of Gizis et al. (2002).  For
the remainder of this paper all results reported will be from spectra that
either passed all of the above criteria (active - 1910 stars) or failed 
all of our tests, including the tests with less stringent criteria 
(inactive - 5930 stars), for a total of 7840 stars in the final activity
sample.  Subdwarfs have also been removed from this sample; they are discussed
separately in $\S$4.4 below.
We reiterate that in order to be classified 
as inactive, the H$\alpha$ EW must be less than 1.0\AA. Thus a star with
H$\alpha$ EW $>$ 1.0\AA, but with noisy data, is not classified as either 
active or inactive, but is simply removed from the sample.
The H02 subset of 659 late type stars from the SDSS
Early Data Release was analyzed with this method and then
inspected by eye.  We found that less than 4\% of the stars were incorrectly
labeled as active or inactive, and cite this as a measure of our ability to 
determine the activity status automatically.

\subsubsection{Quantifying Activity} 

A star's magnetic activity can be 
qualitatively assessed as either active or inactive
using the EW of the H$\alpha$ emission line.
However, because the EW measure depends on the flux in the local stellar
continuum, it cannot be used to compare the strength of the activity in
stars of different spectral type (Reid, Hawley, \& Mateo 1995b). Instead,
we measure activity strength by computing the ratio of the luminosity in H$\alpha$  
(L$_{\rm{H}_\alpha}$) to the bolometric luminosity (L$_{bol}$). 
The ratio L$_{\rm{H}_\alpha}$/L$_{bol}$ is a measure of the magnetically driven
radiative output compared to the total radiative output of the star,
and does not depend on the shape of the stellar continuum.

In the past, L$_{\rm{H}\alpha}$ has been
calculated by converting the measured H$\alpha$ flux 
to a luminosity using the distance to the star. However,
distance determinations for the stars in our sample rest on
either spectroscopic or photometric parallax methods.  
To avoid introducing additional uncertainty into our
L$_{\rm{H}\alpha}$ calculation due to uncertainties in the distances, we
use a distance-independent method of calculating 
L$_{\rm{H}\alpha}$/L$_{bol}$.  The method is described in detail 
in Walkowicz et al. (2004) and uses
well-observed nearby M and L dwarfs with known distances to empirically
determine the ratio of the
luminosity in the continuum at H$\alpha$, L(cont)$_{\rm{H}\alpha}$,
to the bolometric luminosity.  The resulting ratios 
$\chi$ = L(cont)$_{\rm{H}\alpha}$/L$_{bol}$ are tabulated as a function 
of spectral type and color, and when multiplied by the EW in H$\alpha$
for a given star, give the desired ratio L$_{\rm{H}\alpha}$/L$_{bol}$. 

Since the $\chi$ ratios are tabulated at 0.5 spectral type intervals, 
and we use integer spectral types in our analysis, a weighted mean of the 
$\chi$ values surrounding a given integer spectral type was computed.  
We give the two surrounding values 1/4 the weight
of the integer $\chi$ value.  This weight is also used in propagating the
uncertainties.  We multiply each measured H$\alpha$ EW by the appropriate
mean $\chi$ factor and thereby calculate L$_{\rm{H}\alpha}$/L$_{bol}$ 
for each active star in
our sample.  Note that because there are no M0 stars in the nearby star
sample, the M0 L$_{\rm{H}\alpha}$/L$_{bol}$ values in the
present work have been calculated using only the M0.5 $\chi$ value.

\subsection{Colors}
The colors used in our analysis are formed from the $ugriz$ psf 
magnitudes in the SDSS photometric database.  As described above,
we use photometry only from the latest (Fall 2003, Photo v. 5\_4\_25) 
pipeline reductions.
To assure reasonable photometric accuracy, we
include only stars with $u$ magnitude uncertainties less than 0.3 and 
$griz$ magnitude uncertainties less than 0.2 in our analysis of color effects.  This decreases the number of stars available for analysis to $\sim$2200.  However, if the $u$ band data are not required, for example when examining $(g-r)$ colors, the sample numbers $\sim$5800.  The maximum uncertainties
in the colors that we use are therefore 0.36 magnitudes in $(u-g)$ and 0.28 magnitudes
in $(g-r)$, $(r-i)$, and $(i-z)$.  The great majority of stars in this color sample have 
uncertainties much smaller than these maximum values, with typical values
being $\sigma(r)$ $<$ 0.02 for $r < 17$ (Ivezi\'c et al. 2003; 11\% of the sample);
$\sigma(r)$ $<$ 0.03 for $r < 19$ (46\% of the sample); and 
$\sigma(r)$ $<$ 0.04 for $r < 21$ (83\% of the sample).

As discussed above, a significant source of scatter in the blue colors 
of the late type stars in our sample was introduced by the presence of 
numerous white dwarf - M dwarf
binary systems.  The spectroscopically-identified pairs were originally 
removed during the spectral typing process.  In addition, we made a color 
cut on our sample, and included only stars whose colors lie outside the 
locus of known white dwarf - M dwarf pairs in the $(g-r)$ vs. $(u-g)$
color-color diagram (Silvestri, Hawley \& Szkody  2003; Smol\v{c}i\'{c} et al. 2004).  In practice
this meant that only stars with $(u-g)$ $>$ 1.8 are included in the sample.
All suspected pairs based on either photometry or spectroscopy were
therefore removed, and are not included in the final spectroscopic sample of 7840 stars.

\subsection{Molecular Band Indices}

To investigate the presence of metal poor stars in our sample, we
use the varying strengths of the TiO and CaH molecular bands in a
low metallicity environment.  We measure the TiO5, CaH1 and CaH2 
molecular band indices following the procedure outlined by Reid et al.(1995a).
Gizis (1997) uses the same indices to define divisions between dwarfs, 
subdwarfs and extreme subdwarfs.  In our study, the following criteria 
are used to restrict the sample of subdwarfs to high probability candidates: 
a) the band indices must be larger than their uncertainties (typical
uncertainties are 0.04 in TiO5; 0.03 in CaH1; and 0.02 in CaH2);
b) the band heights must be larger than both the noise in the continuum 
near the bandhead and the noise in the bottom of the absorption feature; and
c) the computed uncertainties in the band indices
must not overlap the subdwarf or extreme subdwarf boundaries.

\section{Results} 

\subsection{Activity Fraction vs. Spectral Type}

\begin{figure}
\centerline{\psfig{figure=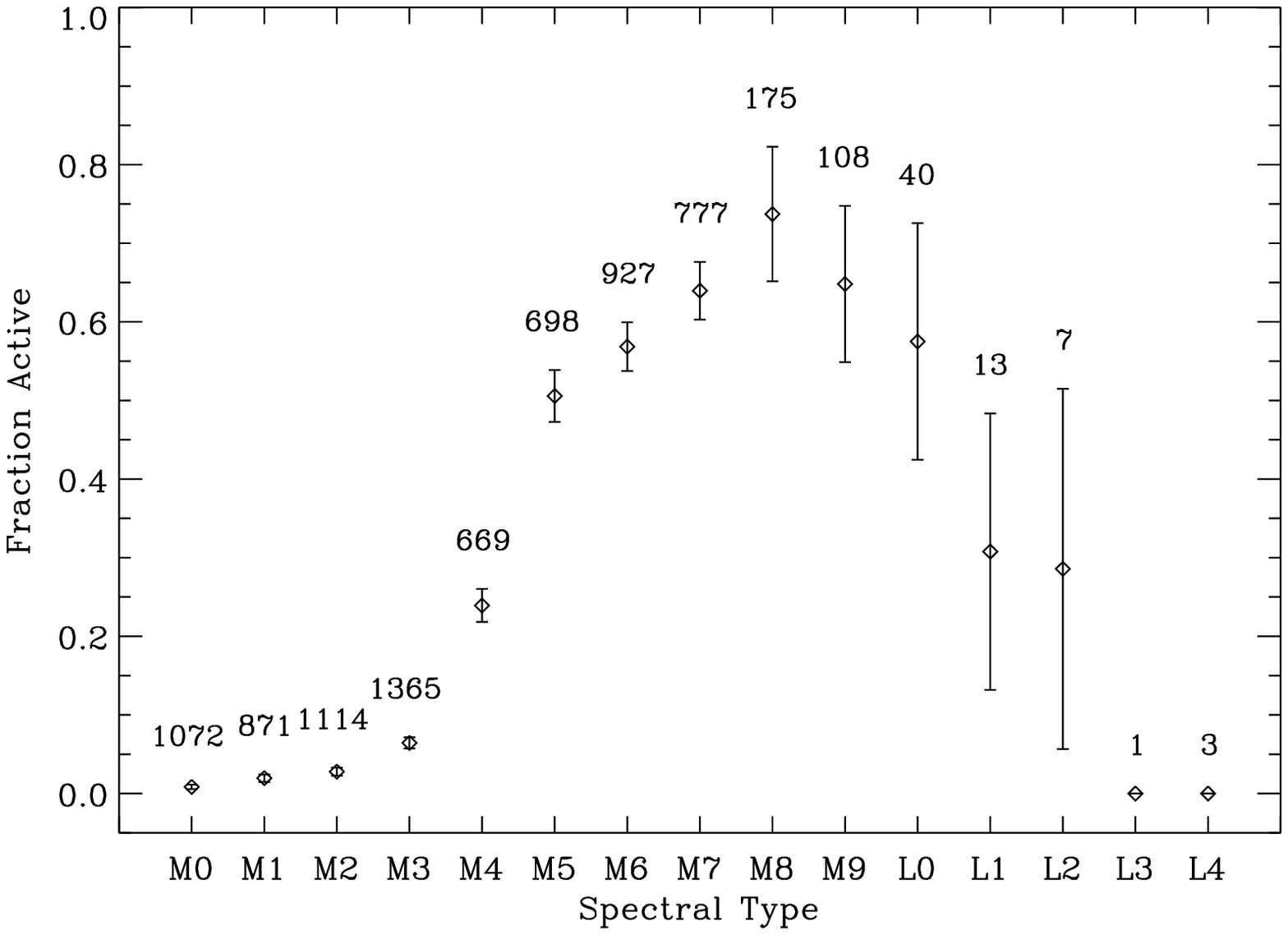,height=12cm}}
\caption{The fraction of active stars is shown as a function of spectral type.  The numbers above each bin represent the total number of stars used to compute
the fraction. The peak at spectral type M8 is consistent with previous results, but the presence of inactive M7 and M8 stars was not seen in earlier 
observations.  Spectral type bins with no observed activity (L3, L4) are 
included for completeness.} 
\label{fig:act} 
\end{figure}

\begin{figure}
\centerline{\psfig{figure=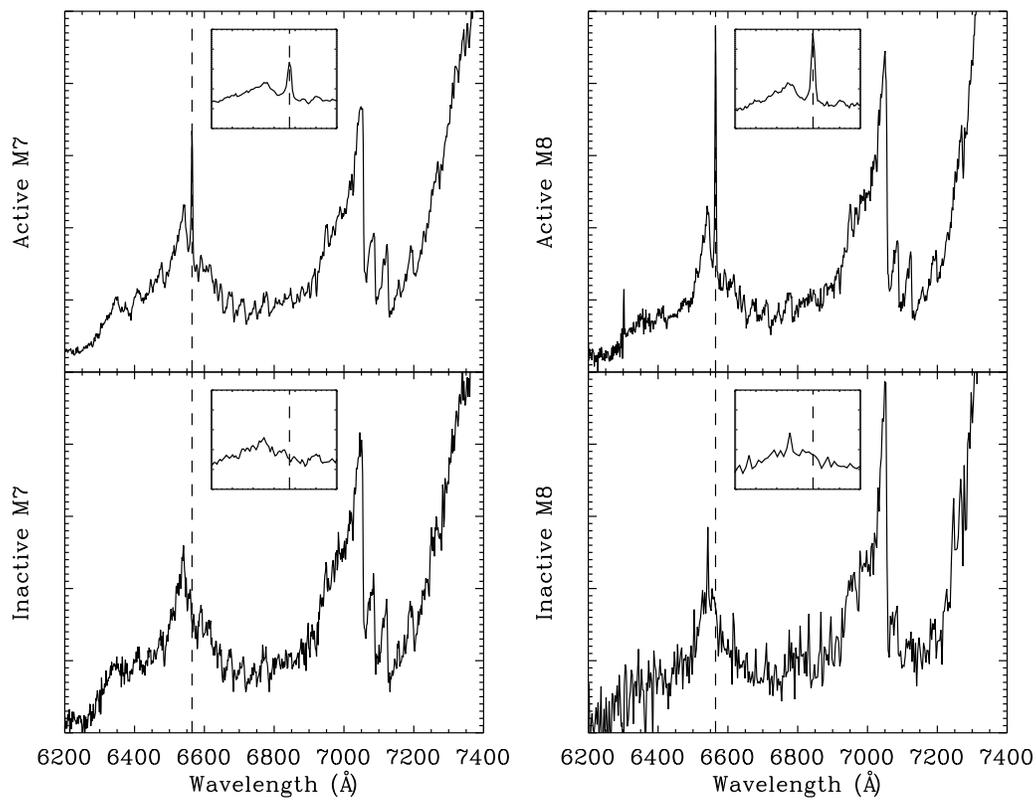,height=12cm}}
\caption{Active and inactive M7 (left panels) and M8 (right panels) stars.  Small boxes in each panel contain a 100\AA\ section of the spectrum near H$\alpha$. 
The dashed 
lines are drawn at the wavelength of H$\alpha$. Note that the line 
falls to the right of the closest spectral feature in both inactive 
spectra. } 
\label{fig:inact} 
\end{figure}

We find that in our sample of 7840 stars
that are either active or inactive, 1910 are classified as active
(24.4\%).  Figure~\ref{fig:act} shows the fraction of stars which are
active as a function of spectral type.  The active fraction peaks at
spectral type M8, where 73\% of stars are active.
% M7,M8,M9 = 64%,73%,65% active 
The numbers above each spectral type bin
represent the total number of stars (both active and inactive) in the bin.
The peak at M8 is consistent with Gizis et al. (2000a), who found 
a peak in the relation at types M7 and M8 using similar activity criteria.
However, all stars in their sample were active at these types, while our
results show $\sim 64-73\%$ of our M7-8 stars are active.
We have confirmed by eye that indeed many of the M7 and M8 
stars in our sample do not have measurable activity.  Figure~\ref{fig:inact}
shows examples of active and inactive M7 and M8 stars.  The continuum
features near H$\alpha$ are noticeable in the inactive stars (blue bump,
red trough), but it is clear that no measurable emission at H$\alpha$
is present.  

\begin{figure}
\centerline{\psfig{figure=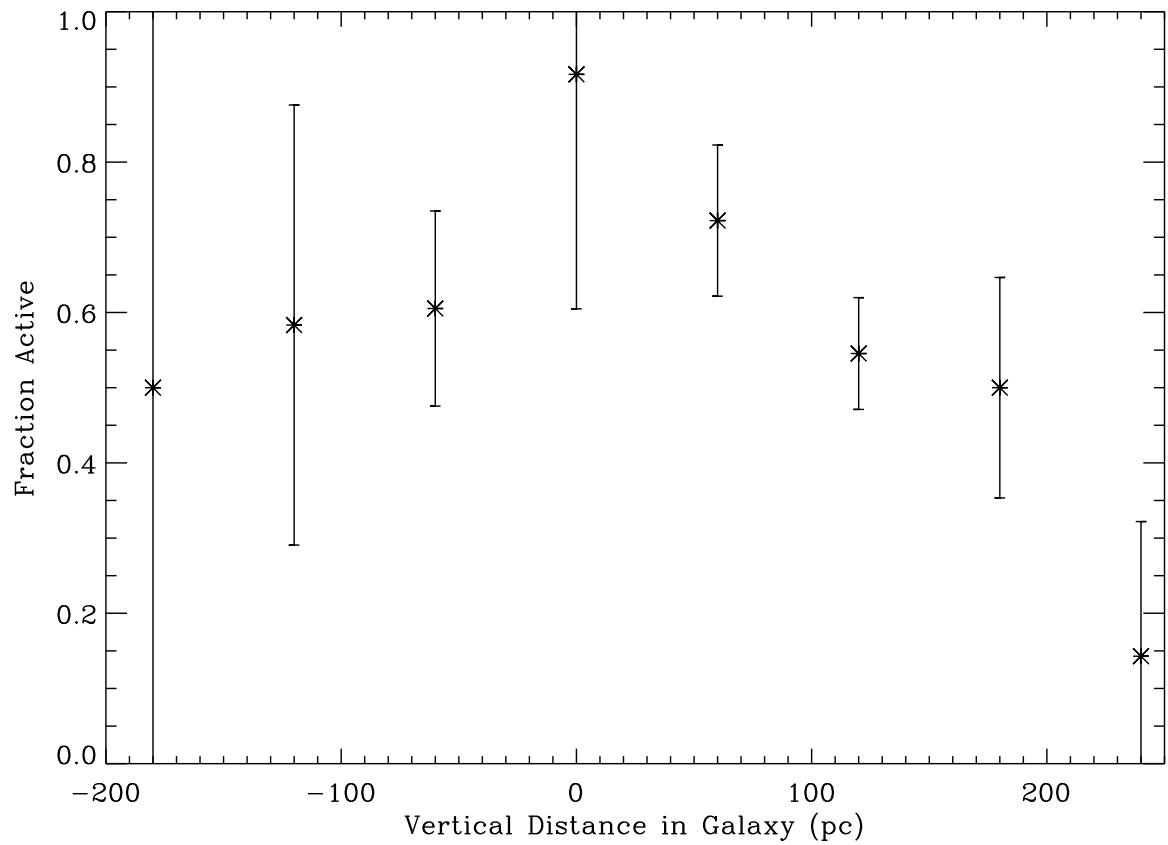,height=12cm}}
\caption{The fraction of M7 stars that are active is shown as a function
of vertical distance above and below the Galactic Plane.  Active stars
appear to be concentrated toward the Plane, which may be an age effect
(see text).}
\label{fig:distance}
\end{figure}

Following a suggestion by Neill Reid (private communication),
we considered whether the disparity in active fraction at type M7 could 
be a distance effect.  The Gizis et al.(2000a) sample was drawn from 2MASS and
represents mostly nearby stars, while the SDSS sample described 
here extends to much greater distances due to the fainter limiting magnitude.
We therefore restricted ourselves to just the M7 stars in the sample,
and computed the distance to each star via the $(i-z)$ photometric parallax 
relation from H02.  Only stars with good photometry ($\sigma(i-z) < 0.28$)
were included (see \S3.3), which reduced the sample from 777 to 269.  Note that we did not
combine stars of different spectral type (color) in the sample, so the 
relative distances between the M7 stars should be quite accurate, as
uncertainties in the absolute calibration of the photometric parallax relation
at different colors do not affect the calculation.
We then computed 
the vertical distance, perpendicular to the Galactic Plane, for each star.
The results are shown in Figure~\ref{fig:distance} for 60 pc distance
bins.  Though the number of stars in each bin is relatively small, leading
to significant Poisson errors, there does appear to be a trend of decreasing 
active fraction with vertical distance from the Plane in both the North
(positive) and South (negative) directions.
If activity is related to age, as discussed by Hawley et al. 
(2000a) and Gizis et al. (2002), then this effect would be a natural consequence
of the increasing velocity dispersion in older disk stars: the older stars 
would be more likely to be found further from the Plane where they were born.
Kirkpatrick et al. (1994) first discussed the possibility that younger stars were
found when looking South (toward the Plane, from the Sun's position 30 parsecs
North of the Plane) than North, out of the Plane.  Small sample sizes and 
uncertain distances precluded a definitive test using data available at that time, but we may now be seeing evidence for this effect in 
our much larger sample.  The uncertainties are still large however, and 
more data are needed to verify this result.  We plan to carry out a rigorous
comparison between the 2MASS and SDSS samples to further investigate the
correlation of activity fraction with distance from the Plane.

\subsection{Activity Strength vs. Spectral Type}
\begin{figure}
\centerline{\psfig{figure=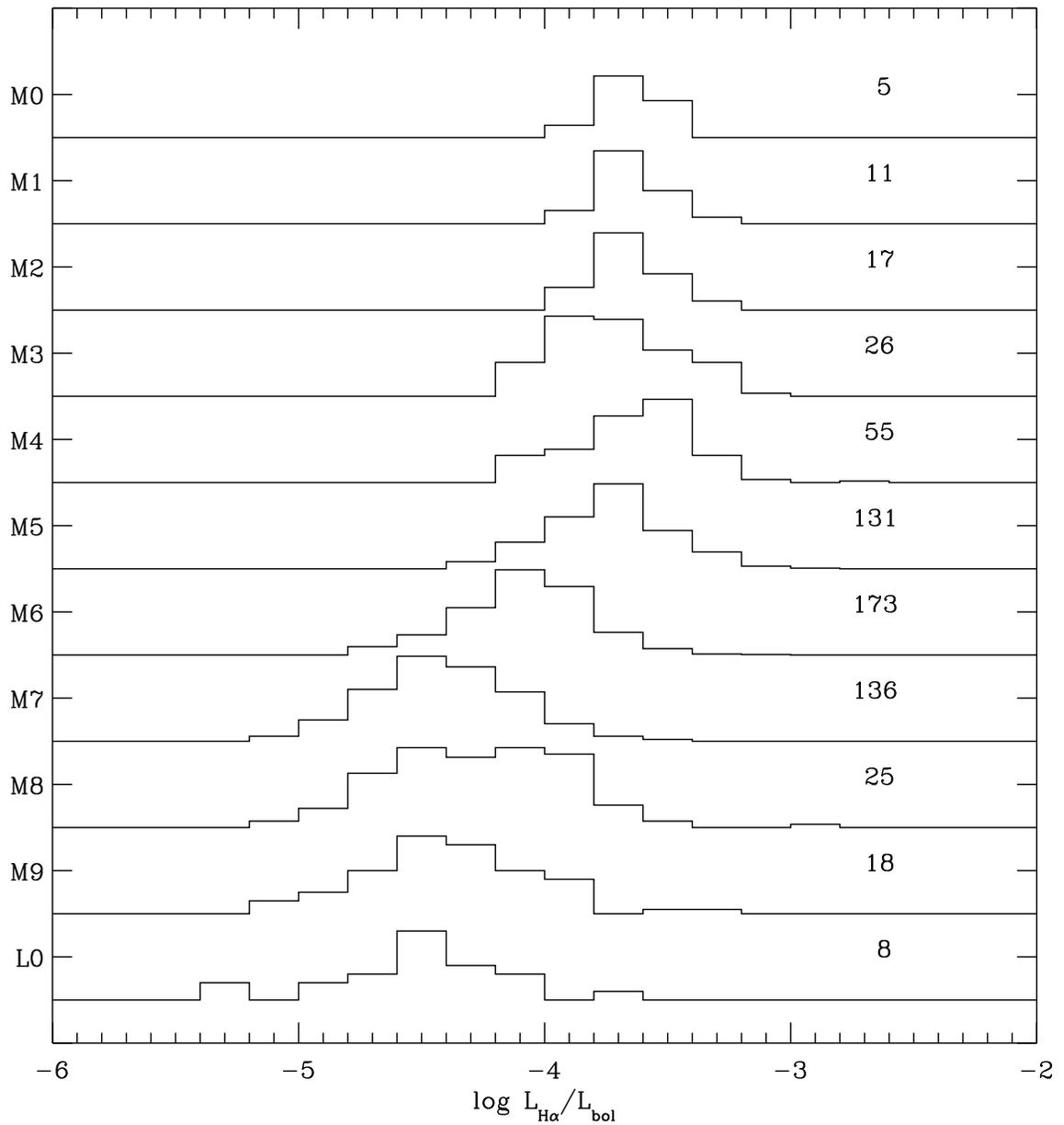,height=17cm}}
\caption{The distribution of the activity strength, 
log(${\rm L}_{\rm{H}\alpha}/{\rm L}_{bol}$), is shown as a function of
spectral type.  The numbers to the right at each spectral type give the peak
value of the histogram.  The mean activity strength clearly declines at types
M6 and later, and the distribution appears to broaden.} 
\label{fig:hist} 
\end{figure}

Figure~\ref{fig:hist} illustrates the activity strength, 
log(${\rm L}_{\rm{H}\alpha}/{\rm L}_{bol}$), 
as a function of spectral type.  Each histogram represents the distribution of
log(${\rm L}_{\rm{H}\alpha}/{\rm L}_{bol}$) 
at a given spectral type.  
The number to the right 
of the  histogram is the peak value of the distribution.  We see that 
log(${\rm L}_{\rm{H}\alpha}/{\rm L}_{bol}$) 
peaks near -3.7 for stars of type M0-M5,
in agreement with Hawley et al. (1996), and systematically peaks at lower
activity strength among stars with later type.
The 1\AA\ H$\alpha$ EW measurement limit in our sample imposes a lower 
limit on the measurable activity strength.  At each spectral type the 
last bin (smallest log(${\rm L}_{\rm{H}\alpha}/{\rm L}_{bol}$)) with stars in it 
corresponds to this measurement limit.  Table~\ref{table:lbolt} gives the log(${\rm L}_{\rm{H}\alpha}/{\rm L}_{bol}$)) corresponding to an H$\alpha$ EW of 1\AA\ for each spectral type.  Thus, the broader distributions seen
at later types may be partly due to a selection effect.  However, the 
distributions among the later types are quite symmetric and definitely
declining at log(${\rm L}_{\rm{H}\alpha}/{\rm L}_{bol}$) values that are easily
measurable.  Therefore it appears that the activity strength covers
a larger range, and is less sharply peaked, among the later type stars.

\begin{figure}
\centerline{\psfig{figure=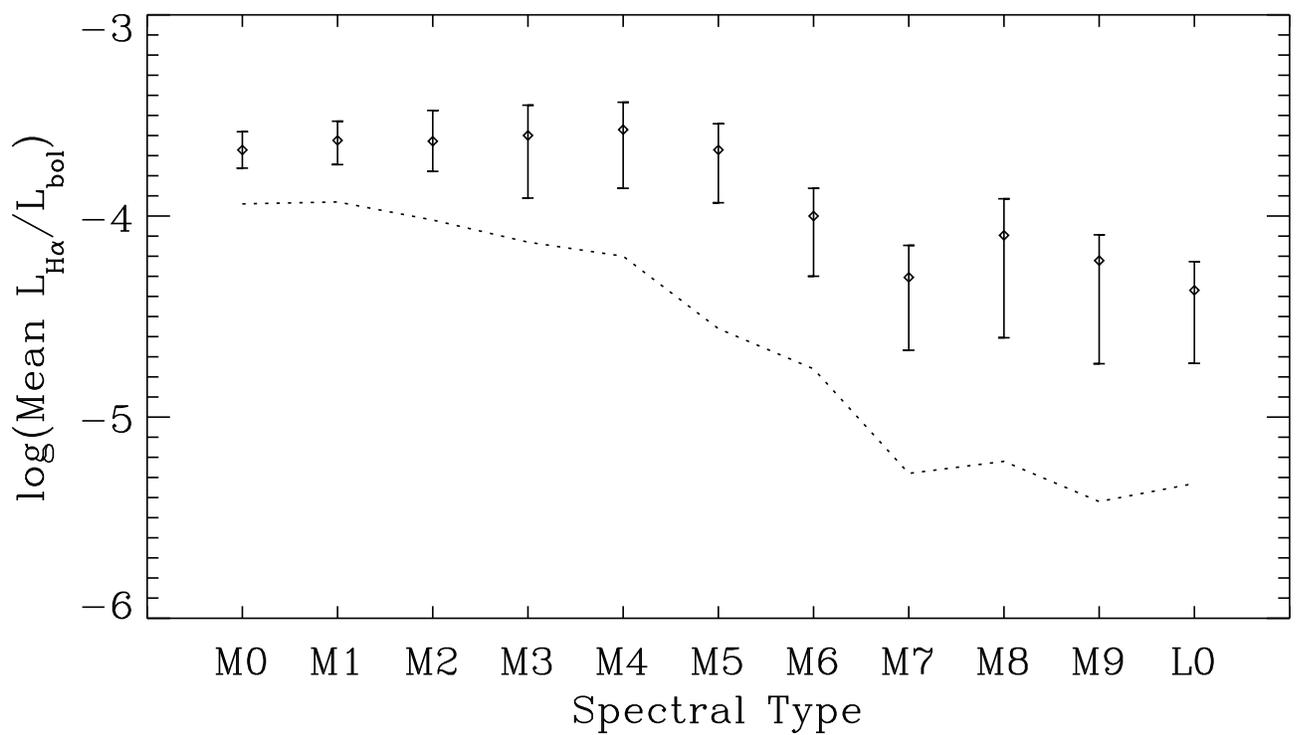,height=11cm}}
\caption{The mean activity strength, log(mean ${\rm L}_{\rm{H}\alpha}/{\rm L}_{bol}$), is
plotted as a function of spectral type.  
Uncertainties reflect the propagated errors in the calculation of
${\rm L}_{\rm{H}\alpha}/{\rm L}_{bol}$ added in quadrature with the 1$\sigma$ distribution
of ${\rm L}_{\rm{H}\alpha}/{\rm L}_{bol}$ in each bin.  The decline in mean activity
strength begins at spectral types M5-M6. The dotted line indicates the ${\rm L}_{\rm{H}\alpha}/{\rm L}_{bol}$ value that we are sensitive to, due to the 1\AA\ EW activity criterion.} 
\label{fig:lbol} 
\end{figure}

Figure~\ref{fig:lbol} shows
the mean of the ${\rm L}_{\rm{H}\alpha}/{\rm L}_{bol}$ distribution as a function of
spectral type.  The errors in this plot are the errors in the
${\rm L}_{\rm{H}\alpha}/{\rm L}_{bol}$ calculation added in quadrature with the 1$\sigma$
distribution of each bin.  The dotted line indicates the ${\rm L}_{\rm{H}\alpha}/{\rm L}_{bol}$ for an H$\alpha$ EW of 1\AA\ (the required minimum EW from $\S$3.2).   Table~\ref{table:lbolt} gives the mean values
for each spectral type.  The approximately constant value from M0-M5 is 
clearly shown,
as is the decline starting at types M5-M6.  This decline occurs at a slightly 
earlier type than in previous studies, where the decline appeared 
at types M7-M8 (Gizis et al. 2000a; Burgasser et al. 2002).  

\subsection{Colors and Activity}

\begin{figure}
\centerline{\psfig{figure=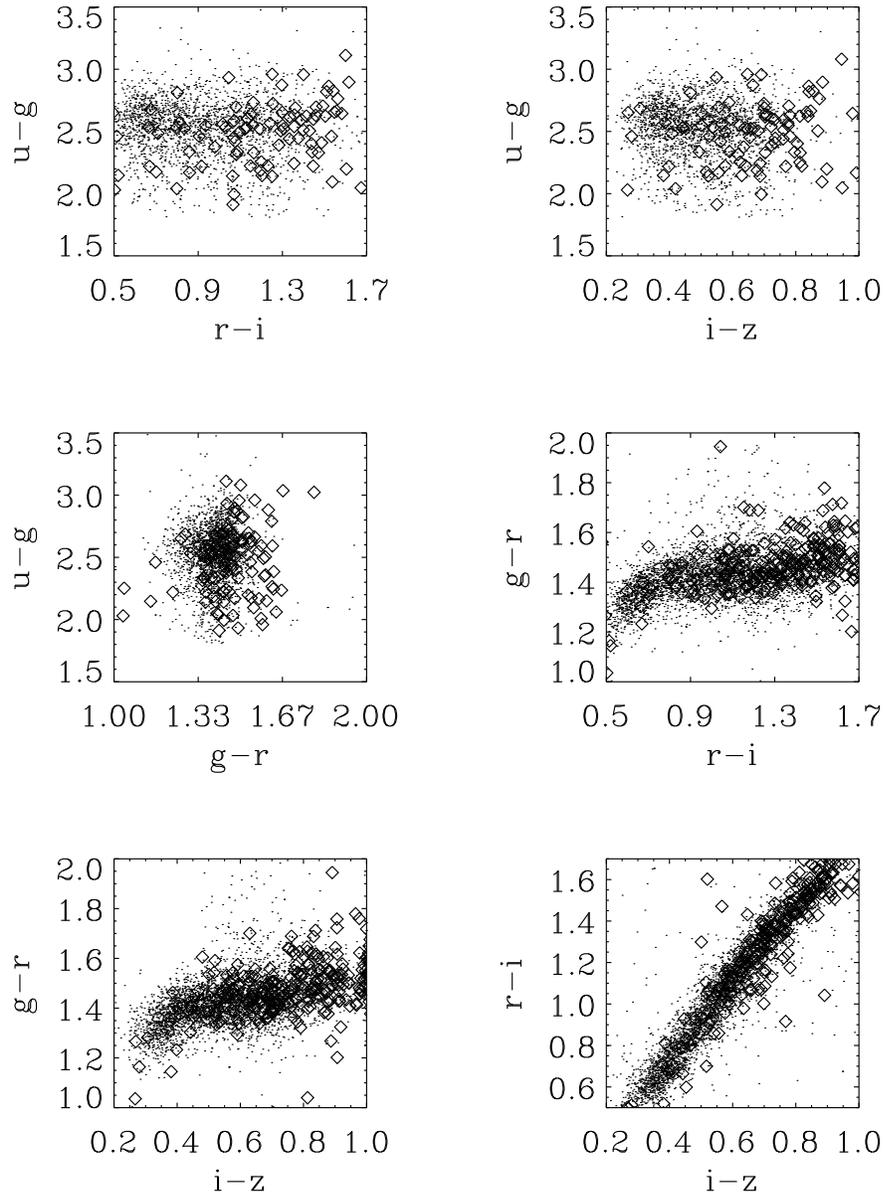,height=17cm}}
\caption{Several color-color diagrams in the SDSS $ugriz$ filters are
used to compare the colors of active stars (diamonds) to inactive 
stars (dots).  A 2D KS test shows that there are no significant differences between active and inactive stars in any of the colors.} 
%maybe change the order of the plots?  seems random at the moment
\label{fig:colors} 
\end{figure}

We used several SDSS color-color diagrams to investigate
the photometric properties of active stars.  Figure~\ref{fig:colors} shows that 
the colors of the active stars are very similar to
the inactive stars.
It has been claimed previously that active stars are marginally bluer than 
inactive ones in the Johnson U-B color, with $\delta$ (U-B) = 0.04 $\pm$ 0.09
for late-type main-sequence dwarfs (Amado \& Byrne 1997).  
They speculate that the effect could be due to
the appearance of bright plage regions on the active stars.  
Our results for the SDSS $(u-g)$ color show a similar, but
not statistically significant effect: $\delta (u-g)$ = 0.05 $\pm$ 0.25.
In the $(g-r)$ color the effect has the opposite sign: $\delta (g-r)$ 
= -0.11 $\pm$ 0.16.  The scatter in the color distribution is much larger 
than the mean offsets, and we therefore cannot confirm the hypothesis that
plage regions are more prevalent on the active stars.  A two-dimensional Kolmogorov-Smirnov test confirms that there is no significant difference between the active and inactive stars in any of the color-color diagrams.

\subsection{Subdwarfs} 
\begin{figure}
\centerline{\psfig{figure=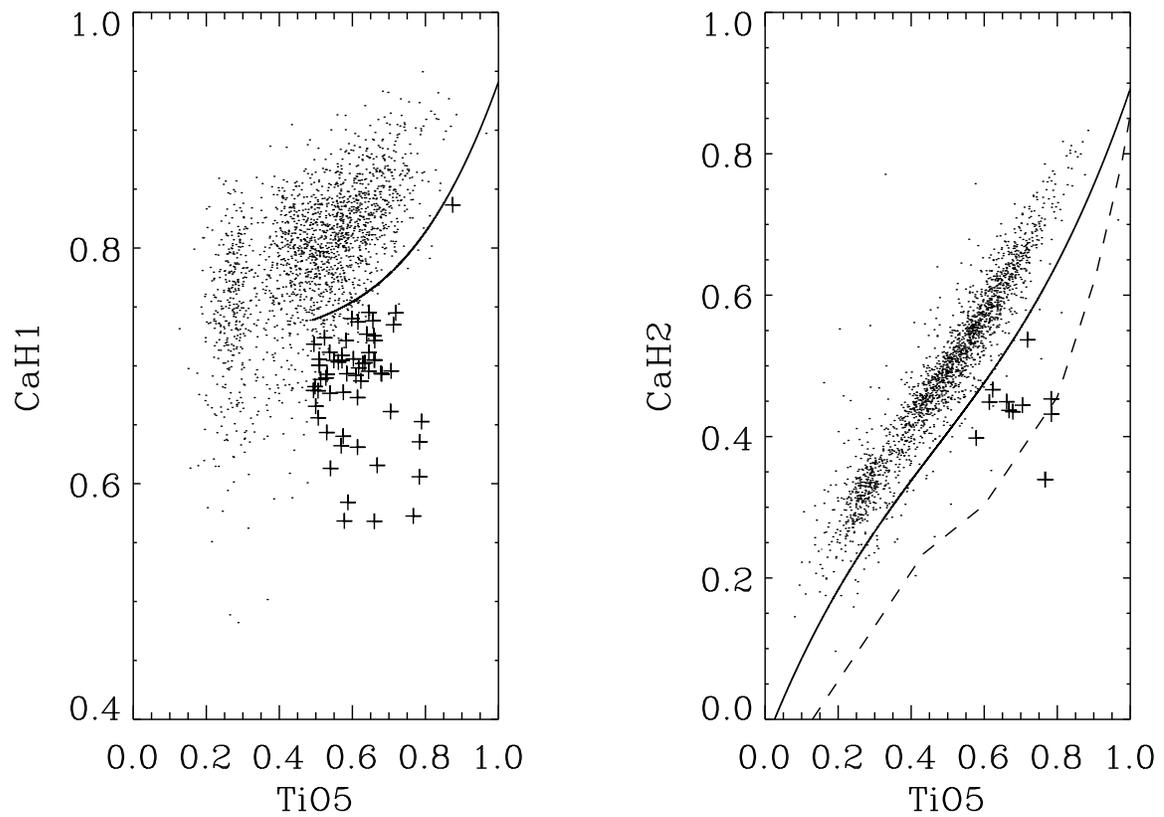,height=12cm}}
\caption{The left panel compares the CaH1 and TiO5 molecular band
strengths, while the right panel compares CaH2 and TiO5.  The solid
line in each panel is the dwarf/subdwarf boundary from Gizis (1997). 
Crosses are subdwarf candidates that meet the criteria defined in $\S$3.4.}  
\label{fig:sub1} 
\end{figure}

%need to redo this plot to stack the panels on top ofeach other instead of
%side by side.  Also need a solid line in cah1 plot.
Figure~\ref{fig:sub1} (left panel)
plots the CaH1 index versus the TiO5 index.  The solid line denotes the
dwarf/subdwarf boundary defined by Gizis (1997).  The CaH1-TiO5
relation is valid only for TiO5 $>$ 0.49 and therefore we have not extended
the boundary beyond that value.  We find 60 stars (crosses) that fall
%is 60 right?
below the subdwarf boundary (and have uncertainties that do not cross the
boundary as described above).  Figure~\ref{fig:sub1} (right panel) shows
a similar plot for the CaH2 index.
Again the solid line indicates the dwarf/subdwarf
boundary and the dashed line is the boundary between the subdwarf and extreme
subdwarf populations.  Several of the stars are close to the extreme 
subdwarf boundary and warrant further observation.  The subdwarf and
extreme subdwarf candidates from this study 
are listed in Table~\ref{table:subdwarf}.
%is the conclusion that only stars that appear as crosses in both plots
%are high prob candidates?  ie there are a lot of stars that look like
%subdwarfs from cah1 but don't in cah2?  are there crosses in the cah2
%plot that are not in cah1?

\begin{figure}
\centerline{\psfig{figure=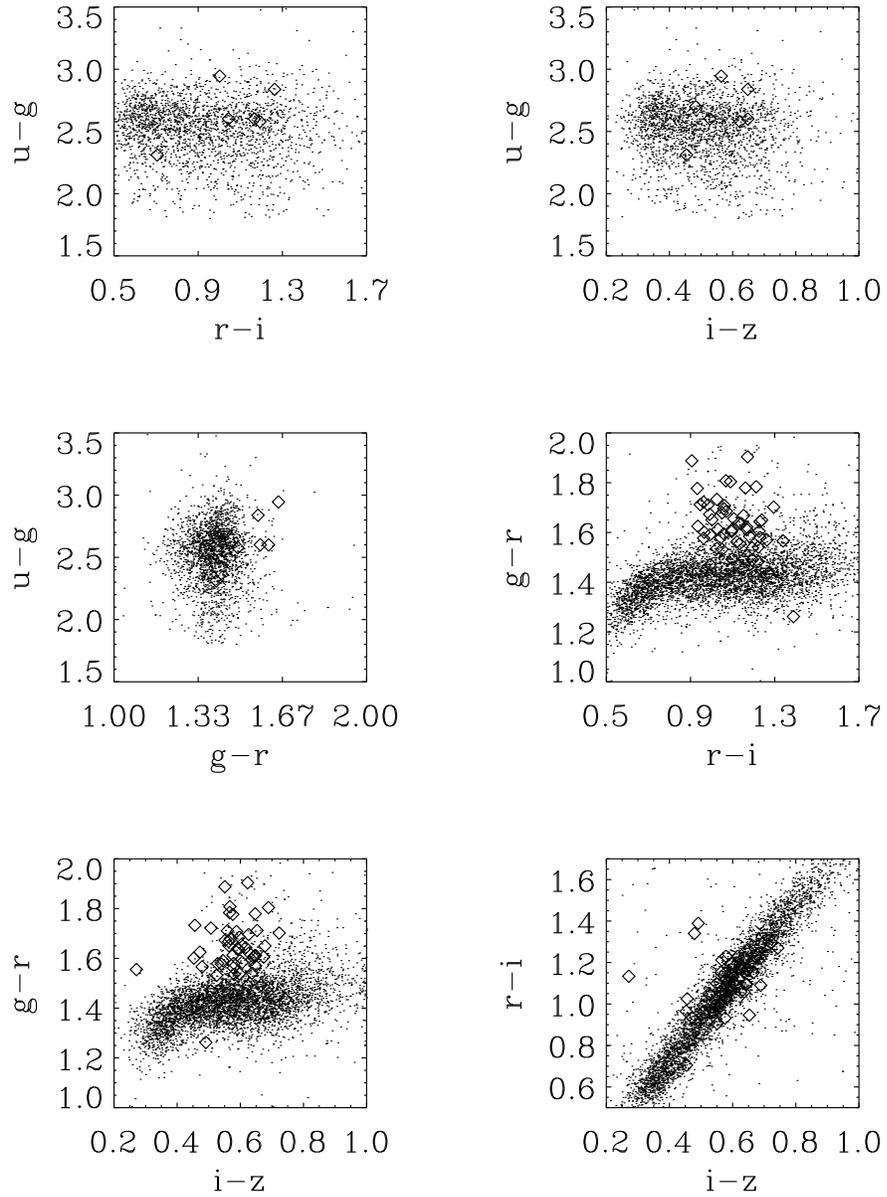,height=17cm}}
\caption{These SDSS color-color diagrams compare the colors of 
subdwarf candidates (diamonds) with normal disk dwarfs 
(dots).  The $(g-r)$ colors of the subdwarf candidates are systematically redder
at a given $(r-i)$ and $(i-z)$ color.} 
\label{fig:colorsub} 
\end{figure}

\begin{figure}
\centerline{\psfig{figure=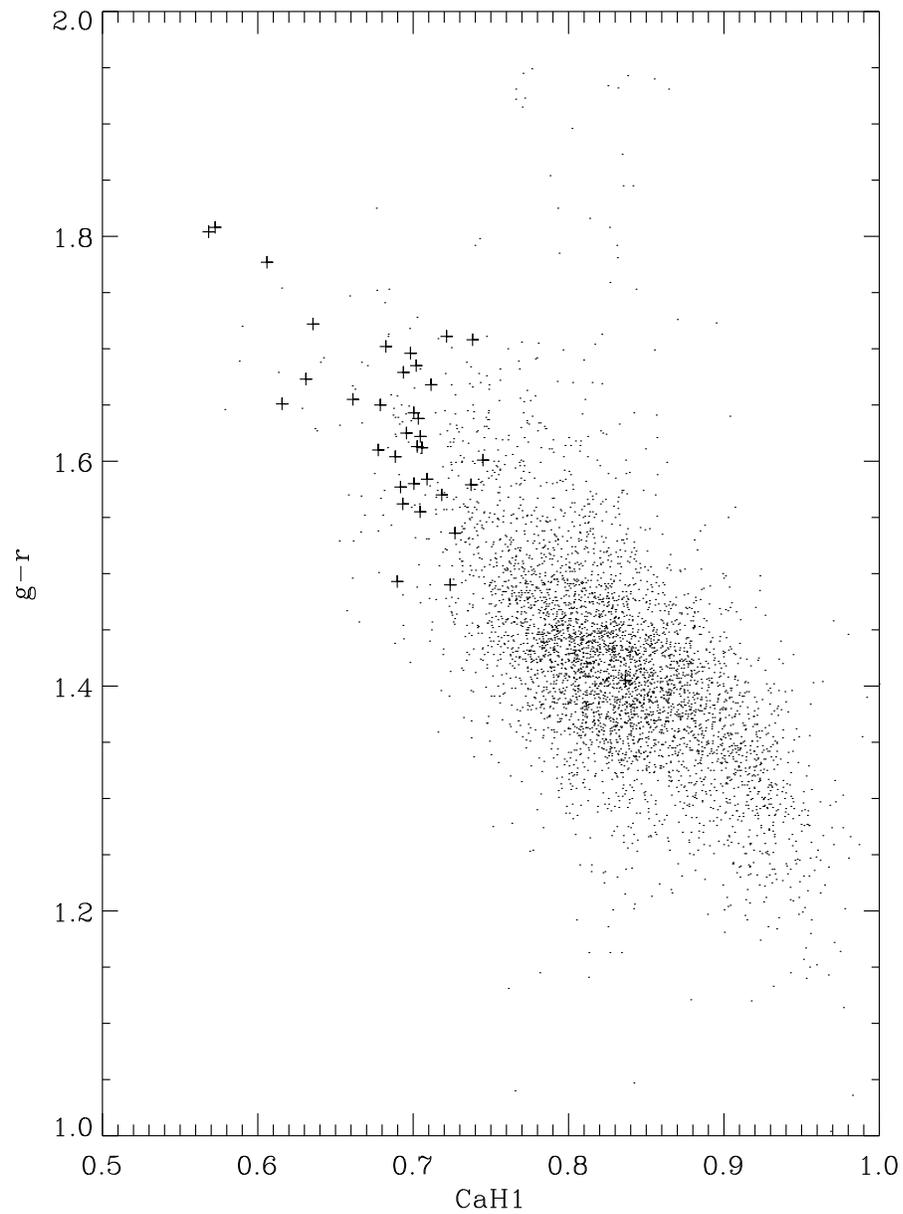,height=17cm}}
\caption{The redder $(g-r)$ colors of the subdwarf candidates are clearly evident in the $(g-r)$ vs. CaH1 diagram.  The SDSS $(g-r)$ color
provides a good photometric proxy for measuring strong hydride bands
and allows subdwarfs to be identified photometrically. To reduce the photometric scatter, only stars with $r<$ 19.5 are shown.} 
\label{fig:colorband} 
\end{figure}

The colors of the subdwarf candidates are compared to the normal disk
population in Figure~\ref{fig:colorsub}.
The subdwarfs are clearly delineated in the $(g-r)$ color, with
the median $(g-r)$=1.61, nearly 0.2 magnitudes redder than the
normal disk dwarfs.  This effect may be due
to excess absorption in numerous hydride bands
in the $g$ bandpass, and has been seen previously
in the Johnson (B-V) color (Hartwick 1977; Bergbusch \& VandenBerg 1992; Gizis 1997; Dahn et al. 1995).
Figure~\ref{fig:colorband} shows the strong correlation between
$(g-r)$ color and CaH1 band strength for stars with $r<$ 19.5.  The subdwarf candidates
are clearly separated from the normal disk dwarfs in this diagram.
This allows subdwarf candidates to be chosen photometrically,
for example in the SDSS spectral targeting
pipeline, with reasonable success.  Note that not all of the stars found here come from subdwarf targeting; some were serendipitous observations
of objects targeted for other purposes.
%true?

\section{Summary}

Using a sample of more than 22000 candidates from the SDSS spectroscopic
database, 7840 late type stars were identified and used to investigate
the magnetic activity in M and early L dwarfs.  This sample is much
larger than all previous samples used for this purpose.  Our results
show that:
\begin{enumerate}

\item{The fraction of active stars rises monotonically from spectral type
M0 to M8 and then declines monotonically to the latest types we
measured (L3-L4).}

\item{Only $\sim 70\%$ of the M7 stars in our sample are active, which
differs from previous studies showing 100\% activity at spectral type M7.
We suggest that the answer to this discrepancy may lie in the distance
distributions of the samples used.  Our results indicate the possibility
that stars near the Galactic Plane are more likely to be active, as
expected if activity depends on age.}

\item{The activity strength, as measured by ${\rm L}_{\rm{H}\alpha}/{\rm L}_{bol}$,
is approximately constant from types M0-M5, and declines at later types. 
The width of the distribution is relatively narrow at early types
but broadens considerably at type M6 and later.}

\item{The average $(u-g)$ color is slightly bluer for active stars compared
to inactive ones, while the average $(g-r)$ color is slightly redder.  
However, these differences are not statistically significant.} 

\item{60 new subdwarf candidates are identified spectroscopically by 
comparing the strengths of the TiO and CaH molecular bands.  Several
of the candidates lie near the extreme subdwarf boundary and may be
interesting targets for future observations.}

\item{The subdwarf candidates have similar colors to the disk dwarfs
except in the $(g-r)$ color, where the median subdwarf $(g-r)$ is almost 0.2
magnitudes redder than disk dwarfs at the same $(r-i)$ color.  This
effect allows subdwarfs to be chosen photometrically from the SDSS
database.}

\end{enumerate}

\section{Acknowledgments}

AAW acknowledges funding from the Astronaut Scholarship Foundation, and the support of Julianne Dalcanton and CWF.

SLH and NMS acknowledge funding from the NSF under grant AST 02-05875.

Funding for the Sloan Digital Sky Survey (SDSS) has been provided by the Alfred P. Sloan Foundation, the Participating Institutions, the National Aeronautics and Space Administration, the National Science Foundation, the U.S. Department of Energy, the Japanese Monbukagakusho, and the Max Planck Society. The SDSS Web site is http://www.sdss.org/.\\

The SDSS is managed by the Astrophysical Research Consortium (ARC) for the Participating Institutions. The Participating Institutions are The University of Chicago, Fermilab, the Institute for Advanced Study, the Japan Participation Group, The Johns Hopkins University, Los Alamos National Laboratory, the Max-Planck-Institute for Astronomy (MPIA), the Max-Planck-Institute for Astrophysics (MPA), New Mexico State University, University of Pittsburgh, Princeton University, the United States Naval Observatory, and the University of Washington.

\clearpage

\clearpage

% we could put the halpha ew=1 limits in this table pretty easily..
\begin{deluxetable}{ccc}
\tablewidth{0pt}

\tablecaption{${\rm L}_{\rm{H}\alpha}/{\rm L}_{bol}$ vs. Spectral Type}

\renewcommand{\arraystretch}{.8}
\tablehead{
\colhead{Spectral Class}&
\colhead{log(mean ${\rm L}_{\rm{H}\alpha}/{\rm L}_{bol}$)}&
\colhead{log(${\rm L}_{\rm{H}\alpha}/{\rm L}_{bol}$) for 1\AA\ EW}}

\startdata

M0& -3.67 & -3.94\\
M1& -3.62 & -3.93\\
M2& -3.63 & -4.02\\
M3& -3.60 & -4.13\\
M4& -3.57 & -4.20\\
M5& -3.67 & -4.56\\
M6& -4.00 & -4.76\\
M7& -4.31 & -5.28\\
M8& -4.10 & -5.22\\
M9& -4.22 & -5.42\\
L0& -4.37 & -5.33\\

\enddata
\label{table:lbolt}
\end{deluxetable}

\clearpage

%\scriptsize
% are the extreme subdwarf candidates marked in here yet?  is the bad one taken
%out?

\begin{deluxetable}{cccccccccc}
\tablewidth{0pt}
\tablecolumns{11} 
%\tabletypesize{\scriptsize}

\tablecaption{Subdwarf Data}

\renewcommand{\arraystretch}{.8}
\tablehead{
\colhead{Name}&
\colhead{$r$\tablenotemark{ab}}&
\colhead{$g-r$\tablenotemark{ab\ }}&
\colhead{$r-i$\tablenotemark{ab\ }}&
\colhead{$i-z$\tablenotemark{b}}&
\colhead{TiO5}&
\colhead{CaH1}&
\colhead{CaH2}&
\colhead{H$\alpha$ EW (\AA)\tablenotemark{c}}&
\colhead{Sp. Type} }

\startdata
SDSS J002228.00$-$091444.8 & 18.92 & -0.89 & 3.44 & 0.48 & 0.65 & 0.71 & 0.52 & 1.16 & 1\nl
SDSS J003541.84$+$003210.1 & 20.15 & 1.64 & 1.14 & 0.59 & 0.68 & 0.69 & 0.54 & -0.16 & 3\nl
SDSS J003701.37$-$003248.3 & 20.15 & 1.60 & 1.23 & 0.67 & 0.57 & 0.64 & 0.46 & -0.23 & 3\nl
SDSS J003755.20$-$002134.2 & 18.70 & --- & 1.01 & 0.89 & 0.65 & 0.75 & 0.57 & -0.35 & 1\nl
SDSS J010811.89$+$003042.4 & 17.34 & 1.61 & 1.17 & 0.65 & 0.51 & 0.71 & 0.45 & -0.43 & 3\nl
SDSS J012853.17$+$003356.6 & 19.34 & 1.56 & 1.12 & 0.58 & 0.58 & 0.69 & 0.50 & -0.21 & 3\nl
SDSS J014631.67$+$001658.3 & 19.79 & 1.26 & 1.39 & 0.49 & 0.58 & 0.72 & 0.48 & -0.18 & 3\nl
SDSS J020753.57$-$001958.8 & 20.16 & 1.54 & 1.22 & 0.64 & 0.50 & 0.67 & 0.46 & 0.02 & 3\nl
SDSS J021956.49$+$005153.6 & 19.27 & 1.60 & 0.97 & 0.45 & 0.72 & 0.74 & 0.54 & 2.97 & 2\nl
SDSS J022221.54$+$005430.3 & 19.01 & 1.64 & 1.13 & 0.59 & 0.56 & 0.70 & 0.46 & 0.04 & 3\nl
SDSS J024501.77$+$003315.8 & 19.38 & 1.54 & 1.03 & 0.58 & 0.64 & 0.73 & 0.51 & --- & 1\nl
SDSS J031314.28$-$000619.8 & 20.33 & 1.48 & 1.22 & 0.69 & 0.51 & 0.66 & 0.51 & -3.27 & 2\nl
SDSS J032749.80$-$004445.4 & 19.88 & 1.59 & 0.99 & 0.55 & 0.64 & 0.70 & 0.62 & -0.35 & 2\nl
SDSS J033408.64$-$072349.2 & 20.21 & 1.89 & 0.91 & 0.55 & 0.79 & 0.65 & 0.45 & -0.85 & 2\nl
SDSS J081329.95$+$443945.6 & 19.39 & 1.68 & 1.07 & 0.60 & 0.63 & 0.70 & 0.50 & -0.45 & 2\nl
SDSS J082230.00$+$471645.8 & 19.39 & 1.67 & 1.15 & 0.56 & 0.54 & 0.71 & 0.46 & 0.04 & 2\nl
SDSS J083002.73$+$483251.6 & 19.92 & 1.90 & 1.17 & 0.62 & 0.53 & 0.69 & 0.45 & 1.89 & 3\nl
SDSS J083217.77$+$522408.2 & 19.61 & 1.49 & 1.10 & 0.55 & 0.60 & 0.71 & 0.52 & -0.38 & 2\nl
SDSS J084105.39$+$032109.6 & 20.05 & 1.45 & 1.15 & 0.58 & 0.62 & 0.69 & 0.47 & 0.24 & 3\nl
SDSS J085843.89$+$511210.1 & 20.00 & 1.53 & 1.05 & 0.52 & 0.66 & 0.70 & 0.60 & -0.98 & 2\nl
SDSS J090238.75$+$471813.6 & 19.97 & 1.73 & 1.02 & 0.46 & 0.66 & 0.71 & 0.64 & -0.58 & 2\nl
SDSS J090434.02$+$513153.9 & 19.40 & 1.78 & 0.93 & 0.58 & 0.78 & 0.61 & 0.45 & -0.64 & 1\nl
SDSS J091451.98$+$453152.8 & 19.02 & 1.64 & 1.23 & 0.62 & 0.51 & 0.70 & 0.50 & -0.58 & 3\nl
SDSS J092429.76$+$523410.7 & 18.70 & 1.67 & 0.99 & 0.55 & 0.61 & 0.63 & 0.45 & -0.47 & 2\nl
SDSS J092534.16$+$524442.4 & 19.79 & 1.71 & 0.95 & 0.65 & 0.71 & 0.73 & 0.58 & -0.61 & 3\nl
SDSS J092708.10$+$561648.1 & 19.37 & 1.62 & 0.93 & 0.47 & 0.71 & 0.70 & 0.62 & -0.41 & 1\nl
SDSS J092745.78$+$582122.7 & 20.47 & 1.78 & 1.21 & 0.57 & 0.59 & 0.58 & 0.55 & -0.86 & 3\nl
SDSS J093024.66$+$554447.7 & 19.14 & 1.56 & 1.13 & 0.27 & 0.57 & 0.70 & 0.48 & -0.21 & 3\nl
SDSS J093141.85$+$453914.5 & 19.45 & 1.58 & 0.96 & 0.54 & 0.61 & 0.69 & 0.57 & -0.41 & 2\nl
SDSS J094306.37$+$465701.4 & 19.74 & 1.59 & 1.06 & 0.64 & 0.54 & 0.61 & 0.50 & -1.34 & 3\nl
SDSS J095147.77$+$003612.0 & 18.27 & 1.58 & 1.04 & 0.53 & 0.62 & 0.74 & 0.53 & -0.35 & 2\nl
SDSS J100109.54$+$015450.2 & 19.12 & 1.71 & 0.98 & 0.56 & 0.66 & 0.72 & 0.45 & -0.62 & 2\nl
SDSS J101031.13$+$651327.6 & 19.39 & 1.68 & 1.05 & 0.57 & 0.68 & 0.69 & 0.44 & 2.80 & 1\nl
SDSS J104320.47$+$010439.4 & 19.16 & 1.66 & 1.11 & 0.60 & 0.70 & 0.66 & 0.44 & --- & 2\nl
SDSS J105122.43$+$603844.8 & 17.15 & 1.65 & 1.00 & 0.56 & 0.67 & 0.62 & 0.44 & -0.17 & 2\nl
SDSS J112751.35$-$001246.8 & 20.03 & --- & 1.07 & 0.65 & 0.66 & 0.73 & 0.56 & --- & 1\nl
SDSS J113501.76$+$033720.3 & --- & --- & --- & --- & 0.53 & 0.64 & 0.53 & --- & 3\nl
SDSS J115900.70$+$665214.3 & 19.37 & 1.61 & 1.09 & 0.65 & 0.58 & 0.68 & 0.47 & -0.53 & 3\nl
SDSS J120724.85$+$004346.4 & 18.10 & 1.49 & 1.19 & 0.62 & 0.52 & 0.72 & 0.49 & -0.44 & 3\nl
SDSS J121510.41$+$003342.7 & 18.00 & 1.41 & 0.70 & 0.45 & 0.87 & 0.84 & 0.77 & -0.19 & 0\nl
SDSS J125919.29$-$025402.3 & --- & --- & --- & --- & 0.60 & 0.74 & 0.52 & -0.39 & 2\nl
SDSS J143930.77$+$033317.3 & 19.38 & 1.80 & 1.09 & 0.69 & 0.58 & 0.57 & 0.40 & -0.85 & 3\nl
SDSS J145447.32$+$011006.8 & 20.35 & 1.44 & 1.27 & 0.74 & 0.57 & 0.63 & 0.55 & -0.57 & 3\nl
SDSS J145547.00$+$602837.3 & 19.18 & 1.70 & 1.06 & 0.63 & 0.62 & 0.70 & 0.47 & -0.28 & 2\nl
SDSS J150511.33$+$620926.3 & 18.61 & 1.81 & 1.07 & 0.57 & 0.77 & 0.57 & 0.34 & -0.38 & 3\nl
SDSS J161348.84$+$482016.0 & 18.30 & 1.72 & 0.96 & 0.51 & 0.78 & 0.64 & 0.43 & -0.34 & 1\nl
SDSS J171745.22$+$625337.0 & 18.66 & 1.65 & 1.24 & 0.68 & 0.51 & 0.68 & 0.42 & 0.10 & 3\nl
SDSS J173452.52$+$603603.1 & 18.78 & 1.71 & 1.06 & 0.59 & 0.66 & 0.74 & 0.55 & -0.28 & 2\nl
SDSS J215937.69$+$005536.2 & 18.96 & 1.61 & 1.03 & 0.58 & 0.63 & 0.70 & 0.53 & --- & 2\nl
SDSS J221500.88$+$005217.2 & 19.08 & 1.60 & 1.09 & 0.64 & 0.51 & 0.69 & 0.49 & -0.28 & 2\nl
SDSS J221625.03$-$003122.5 & 19.28 & 1.58 & 1.18 & 0.61 & 0.57 & 0.71 & 0.53 & -0.49 & 3\nl
SDSS J223802.82$-$082532.4 & 19.71 & 1.78 & 1.16 & 0.65 & 0.49 & 0.68 & 0.45 & -0.65 & 3\nl
SDSS J224605.41$+$141640.6 & 17.00 & 1.57 & 1.26 & 0.65 & 0.49 & 0.72 & 0.45 & -0.41 & 3\nl
SDSS J224854.83$-$091723.2 & 19.85 & 1.57 & 1.34 & 0.48 & 0.54 & 0.68 & 0.50 & -0.21 & 3\nl
SDSS J225538.34$-$005945.2 & 19.89 & 1.47 & 0.99 & 0.51 & 0.61 & 0.67 & 0.55 & 0.21 & 1\nl
SDSS J230303.49$-$010656.7 & 18.95 & 1.62 & 1.17 & 0.63 & 0.55 & 0.70 & 0.50 & -0.40 & 3\nl
SDSS J230805.24$+$001812.7 & 19.36 & 1.70 & 1.30 & 0.72 & 0.50 & 0.68 & 0.42 & -0.49 & 3\nl
SDSS J233030.19$+$004521.9 & 19.45 & 1.58 & 1.23 & 0.58 & 0.51 & 0.70 & 0.46 & -0.80 & 3\nl
SDSS J235116.25$-$003104.8 & 19.48 & 1.49 & 1.15 & 0.67 & 0.53 & 0.69 & 0.45 & -1.30 & 3\nl
SDSS J235830.60$-$011413.2 & 19.99 & 1.54 & 1.19 & 0.60 & 0.66 & 0.57 & 0.49 & -0.61 & 2\nl
\enddata

\tablenotetext{a}{Photometry is omitted when uncertainties are larger than 0.2 magnitudes.}
\tablenotetext{b}{Photometry is not available for some objects.}
\tablenotetext{c}{EWs of subdwarfs that were either classified as active or inactive (see Section 3.2) are included.}
\label{table:subdwarf}

\end{deluxetable}

\end{document}